\documentclass[12pt]{article} \usepackage{epsfig,amsmath,graphicx,amssymb}
\begin{document}
\titlepage

\begin{flushright}
OCHA-PP-315\\   TU-936\\   May 2013\\ revised: August 2013\\ 
\end{flushright}

\vspace*{1.0cm}

\begin{center}
{\Large \bf 
Constraints on radion in a warped extra dimension model from 
Higgs boson searches at the LHC
} 

\vspace*{1.0cm} 

Gi-Chol Cho$^a$, Daisuke Nomura$^b$ and Yoshiko Ohno$^c$

\vspace*{0.5cm}
$^a${\em Department of Physics, Ochanomizu University,
    Tokyo 112-8610, Japan}\\
$^b${\em Department of Physics, Tohoku University, Sendai 980-8578, Japan}\\
$^c${\em Graduate School of Humanities and Sciences, 
         Ochanomizu University, Tokyo 112-8610, Japan}\\
\end{center}

\vspace*{1cm}

\begin{abstract}
\noindent
We study constraints on the radion mass and couplings in the
 Randall-Sundrum model from the recent LHC data 
on the Standard Model (SM) Higgs boson searches.  
When the radion is heavy enough so that it can decay into a pair of
on-shell $Z$-bosons, we find that the $ZZ$ channel gives a stringent  
constraint. 
For example, if the radion mass $m_\phi$ is 200 GeV,
the scale $\Lambda_\phi$ which characterizes the
interactions of the radion with the SM fields must be larger than 5 TeV.
Even for $m_\phi=1$ TeV, we find that the lower bound
on $\Lambda_\phi$ is 2 TeV.
\end{abstract}

\section{Introduction}
The Randall-Sundrum (RS) model~\cite{Randall:1999ee} is an
attractive new physics scenario beyond the Standard Model (SM).
In the RS model, we assume that there exists a warped extra dimension 
compactified on an orbifold $S^1/Z_2$, and that on each boundary
of the fifth dimension, there exists a 3-brane.
The metric of the five dimensional spacetime in this model
is non-factorizable and given by
\begin{align}
 ds^2 = e^{- 2 k y} \eta_{\mu\nu} dx^\mu dx^\nu - dy^2~, 
\end{align}
where $x^\mu$ $(\mu = 0, \cdots, 3)$ are the coordinate
of four-dimensional spacetime, $y$ is the coordinate
of the extra dimension, $0\le y \le \pi r_c$, and $k$ is
of the order of the Planck scale.  
The flat Minkowskian metric $\eta_{\mu\nu}$ is taken to be 
$\eta_{\mu\nu} = {\rm diag}(+1, -1, -1, -1)$.   
The brane on which we
live, called visible (or TeV) brane, is located at $y=\pi r_c$,
and the other brane, called hidden (or Planck) brane, is located
at $y=0$.  
Because of the so-called ``warp factor'' $e^{-2k\pi r_c}$ in the metric, the  
hierarchy between the TeV and Planck scales is explained when 
$kr_c \simeq 12$.  

In the original RS model, it was unknown how to fix the
value of $r_c$ as opposed to other values.  
In Ref.~\cite{Goldberger:1999uk} the authors introduce
a bulk scalar into the model, and assume that it has 
self-couplings localized on the branes.  
Remarkably, under this situation, the potential has a minimum at finite
$r_c$, which can be chosen so that $kr_c \sim {\cal O}(10)$.  
The particle associated with the degree-of-freedom for
the choice of the distance between the branes is
called the radion.  

The radion is phenomenologically interesting, since
it is expected to have a mass around the TeV scale.
Moreover, since it comes from the gravity degrees-of-freedom,
its couplings to the SM fields are fixed: it couples to the trace
of the energy-momentum tensor of the SM fields.  Since the
energy-momentum tensor contains quadratic terms in the SM fields,
the radion can decay into a pair of the SM particles like
$W^+W^-, ZZ$ and $\gamma\gamma$.  Since these decay modes are very
similar to those of the SM Higgs boson~\cite{Goldberger:1999un},
we can constrain
the mass and couplings of the radion using the data for Higgs boson
searches at the LHC.  (For earlier works on this line of argument,
see Refs.~\cite{Barger:2011qn, deSandes:2011zs, Cheung:2011nv,
Ohno-Cho}.  
For more general aspects on radion phenomenology at the LHC,
see e.g.\ Refs.~[8--34].)

In this paper, we study the experimental bound on the 
radion parameters (mass and couplings) allowed by the LHC data 
for the SM Higgs boson searches.  In Sections 2 and 3,
we discuss the production of the radion at the LHC
and its decay modes, respectively.
These two sections are preparations for Section 4, where
we give our numerical results on the constraint
on the radion mass and couplings.
Section 5 is devoted to the summary of our work
and discussions.

\section{Production of radion at LHC}

The interaction Lagrangian of the radion $\phi$ to the SM
fields is,
\begin{align}
 {\cal L}_{\rm int} =
 \frac{\phi}{\Lambda_\phi} T^{\mu}_{~\mu}~,
\end{align}
where $T^{\mu}_{~\nu}$ is the energy-momentum tensor of the
SM fields, and $\Lambda_\phi$ is the vacuum expectation value
of the radion, which is expected to be the TeV scale.   
The trace of the energy-momentum tensor reads
\begin{align}
 T^{\mu}_{~\mu} &=  - 2 m_W^2 W_\mu^+ W^{- \mu}
  -   m_Z^2 Z_\mu Z^\mu
 + \sum_f m_f \bar{f} f
 + (2m_h^2 h^2 - \partial_\mu h  \partial^\mu h) + \cdots \nonumber\\
& \quad
 + \sum_{a=1,3} \frac{\beta_a(g_a)}{2g_a}
          F^a_{\mu\nu}F^{a\mu\nu} ~,
\label{eq:trace_SM_EMtensor}
\end{align}
where the sum in the last term (called the anomaly term since it
comes from the trace anomaly) should be taken over QED (corresponding
to $a=1$) and QCD ($a=3$), and the ellipsis means higher order terms.
$h$ is the SM Higgs boson, and $f$ stands for the SM fermions.
To clarify the normalization of $\beta_a$, we note here
that $\beta_3/(2g_3) = - (\alpha_s/8\pi) b_{\rm QCD}$ with
$b_{\rm QCD}=11-2n_f/3$, where $n_f$ is the number of 
active quark-flavors (we take $n_f=6$ throughout this paper).

In general, the radion can mix with the Higgs boson through
the Higgs-curvature mixing term~\cite{Giudice:2000av, Csaki:2000zn}.  
However, in this paper we do not consider the radion-Higgs mixing 
since it is reported that the decay branching ratios of the
Higgs boson discovered at the LHC~\cite{Aad:2012tfa,
Chatrchyan:2012ufa} are consistent with those
of the SM Higgs boson~\cite{Moriond2013Higgstalks}.   

For the radion mass at sub-TeV to TeV scale, the main production 
mechanism of the radion at the LHC is the gluon fusion.
The main difference between the gluon-fusion production of the
SM Higgs boson and that of the radion comes from the anomaly term
in Eq.~(\ref{eq:trace_SM_EMtensor}), which gives 
an important contribution 
in addition to the fermion-loop (mainly from the top-quark loop)
diagrams.  
As a consequence, the production cross section of 
$\phi$ is larger than that of the SM Higgs boson of the same
mass by a factor of $\sim {\cal O}(10)$ for
$\Lambda_\phi=1{\rm TeV}$~\cite{Giudice:2000av}.

\begin{figure}[t]
\begin{center}
\includegraphics[scale=1.0]{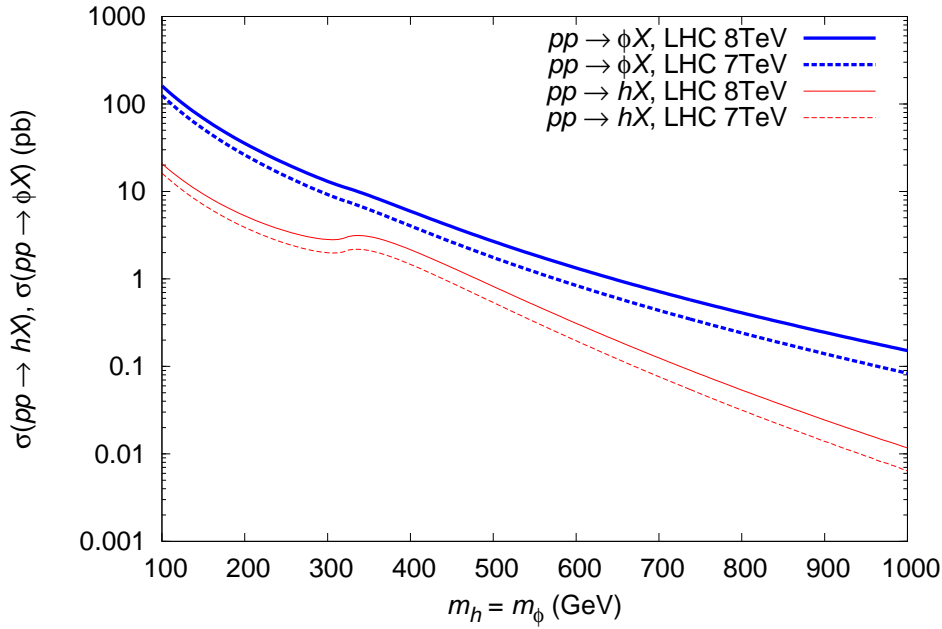}
\end{center}
\vspace*{-1.cm}
\caption{\label{fig:sigma_ppH_ppphi}
The production cross sections of the radion at the LHC
with the center-of-mass energy $\sqrt{s}=8$ TeV (thick solid line)
and 7 TeV (thick dashed line).
Also shown are the production cross sections of the SM Higgs boson
at the LHC with the same beam energies (thin solid line for
$\sqrt{s}=8$ TeV and thin dashed line for $\sqrt{s}=7$ TeV).
The radion interaction scale $\Lambda_\phi$ is taken to be
1 TeV in the figure.}
\end{figure}
The radion production cross sections at the LHC
with the center-of-masses $\sqrt{s} = 7$ TeV and 8 TeV are shown in
Fig.~\ref{fig:sigma_ppH_ppphi} as a function of its mass $m_\phi$. 
In the figure, the radion interaction scale $\Lambda_\phi$
is taken to be 1 TeV.
Also shown are the production cross sections of the SM Higgs
boson of the same mass and at the same LHC beam energies for comparison.
To compute the Higgs boson production cross sections, we first calculate
the leading-order (LO) cross section of $pp \to h X$,
and multiply it by the $K$-factor, which we take to be the central value
of the NNLO $K$-factor given in Fig.~8 of Ref.~\cite{Catani:2003zt}.
To calculate the radion production cross sections, we assume that the
$K$-factor for the radion production is the same as that for the
production of the SM Higgs boson of the same mass, and multiply
the LO radion production cross sections by the same $K$-factor.
To calculate the LO cross sections we use the CTEQ6L parton distribution
functions~\cite{Pumplin:2002vw}.
As already discussed, the radion production cross section is
larger than that of the SM Higgs boson of the same mass
because of the enhancement
factor from the anomaly term in the interaction Lagrangian.

\section{Decays of radion}
\begin{figure}[t]
\begin{center}
\includegraphics[scale=1.0]{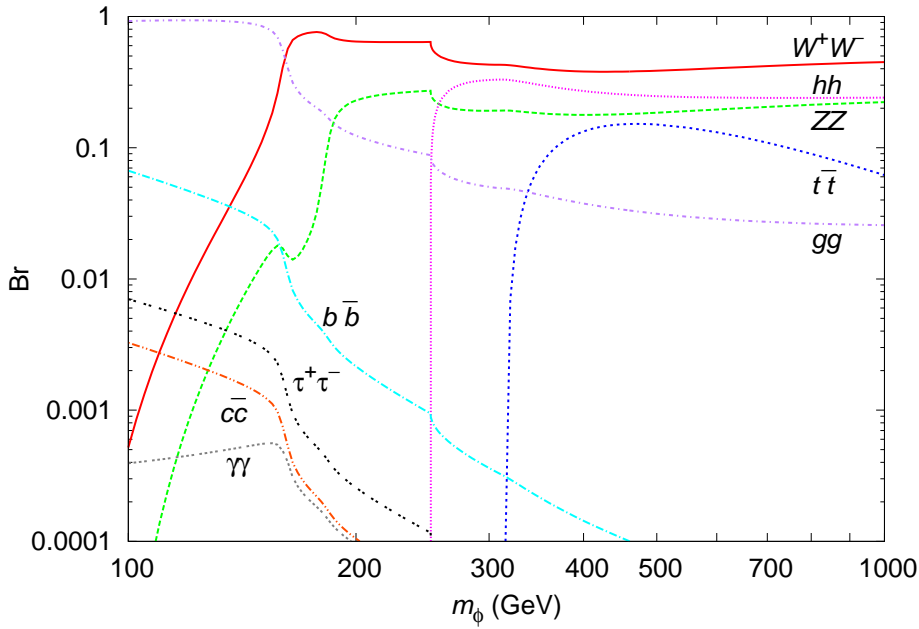}
\end{center}
\vspace*{-1.cm}
\caption{\label{fig:phidecays}
The decay branching ratios of the radion.  The SM Higgs boson
mass is taken to be $125.5$ GeV.}
\end{figure}

The decay branching ratios of the radion are dictated
by the interaction Lagrangian, Eq.~(\ref{eq:trace_SM_EMtensor}).  
In Fig.~\ref{fig:phidecays} we show the behaviors of the
branching ratios as functions of the radion mass.
In the figure, we take the SM Higgs boson mass to be
125.5 GeV~\cite{Aad:2012tfa, Chatrchyan:2012ufa}.

To compute the branching ratios, we have included various 
radiative corrections.
The followings are a list of the corrections:
\begin{itemize}
 \item $gg$ channel

  Following Ref.~\cite{Barger:2011qn},
  we assume that the $K$-factor for the decay rate
  is the same as that of the 
  SM Higgs boson production cross section of the same mass,
  and multiply the LO $gg$ decay rate by
  (the central value of) the NNLO $K$-factor given in Fig.~8 of
  Ref.~\cite{Catani:2003zt}.

 \item $VV$ channels ($V=W, Z$)

  Above the nominal threshold, we use a tree level expression.
  Below the threshold, we include contributions from three-body
  decays ($\phi \to VV^* \to V f\bar{f}$).  We neglect
  four-body decays ($\phi \to V^*V^*$ where both $V$ are off-shell)
  since it is known to be small for our purposes~\cite{Djouadi:2005gi}.
  Very close to the nominal threshold (but below it), we include
  the effect of non-vanishing width of $V$ following Ref.~\cite{Keung:1984hn}.

 \item $q\bar{q}$ channels ($q=t, b, c$)

   We use the NNLO running quark mass to evaluate the decay widths.
   In addition, we apply ${\cal O}(\alpha_s^2)$ QCD
   corrections~\cite{Gorishnii:1990zu}.

\end{itemize}

In the next section, we need the SM Higgs boson decay branching
ratios.  When calculating them, we include similar corrections
as above.

When the radion is heavy enough, the $W^+W^-$ decay mode is the
dominant one, and the branching ratio can be ${\cal O}(1)$.
At the region $m_\phi \gtrsim 200$ GeV, the $ZZ$ decay branching
ratio is about the half of that of $W^+W^-$.  Since these features
are similar to those of the SM Higgs bosons, we can make use of
the $W^+W^-$ and $ZZ$ data for the SM Higgs boson searches
to the case of the radion as well.  

At the low-mass region $m_\phi \sim 120$ GeV, the radion can decay
into a pair of photons mainly via the anomaly term in addition to the
contributions from $W$-boson and top-quark loop diagrams.  Although the
partial decay width to $\gamma\gamma$ is enhanced by the anomaly
term, the full width at this mass region
is dominated by the $gg$ decay width, which is also enhanced by the
contribution from the anomaly term.  As a result, the
branching ratio to $\gamma\gamma$ is suppressed, rather than 
enhanced, compared to that of the SM Higgs boson.

\section{Constraint on radion mass and couplings from LHC data}

With the preparations in the previous sections, we are now
ready to discuss the constraint on the radion mass and couplings
from the Higgs-boson search data at the LHC.

The ATLAS and CMS collaborations recently discovered the SM Higgs
boson at $m_h \simeq 125.5$ GeV~\cite{Aad:2012tfa, Chatrchyan:2012ufa}.  
At the same time, they have
excluded the existence of the SM Higgs boson at wide range of $m_h$
other than $m_h \simeq 125.5$ GeV from non-observation of
$pp\to h \to W^+ W^-, ZZ, \gamma\gamma$.  As already discussed
by a number of authors~\cite{Barger:2011qn, deSandes:2011zs,
Cheung:2011nv, Ohno-Cho},
these data can be used to exclude the radion in the RS
model since its production mechanism and decay modes are
similar to the SM Higgs boson.

In this paper, we use the CMS data for the 
$W^+W^-, ZZ$ and $\gamma\gamma$ decay modes of the SM Higgs
boson~\cite{CMS-PAS-HIG-13-003, CMS-PAS-HIG-13-002, CMS-PAS-HIG-13-001}.
To evaluate the excluded region, we use the following method.
Below we explain it taking the $ZZ$ channel as an example.

In the left panel of Fig.~5 of Ref.~\cite{CMS-PAS-HIG-13-002},
the authors give values of
the 95\% CL limit on $\sigma/\sigma_{\rm SM}$ as a function of
the Higgs boson mass $m_h$, which is equal to the invariant mass of the
$Z$-boson pair in this case.  We call this curve (the solid curve
with the label ``Observed'') $f(m_{h})$.  To interpret this as
a constraint on the radion, we impose the condition below,
\begin{align}
& \left[  \int {\cal L}_{\rm 7{\rm TeV}} dt
            \cdot \sigma(pp \to \phi X; 7 {\rm TeV})
        + \int {\cal L}_{\rm 8{\rm TeV}} dt 
            \cdot \sigma(pp \to \phi X; 8 {\rm TeV})
\right] {\rm Br}(\phi \to ZZ)
\nonumber\\
& \le  f(m_{h}) 
 \left[  \int {\cal L}_{7{\rm TeV}} dt 
              \cdot \sigma(pp \to h X; 7 {\rm TeV})
       + \int {\cal L}_{8{\rm TeV}} dt
              \cdot \sigma(pp \to h X; 8 {\rm TeV})
\right]  \nonumber\\
& \quad 
   \times {\rm Br}(h \to ZZ)  \bigg|_{m_h = m_\phi}~,
\label{eq:the_condition}
\end{align}
where the right-hand side should be evaluated with the understanding 
that $m_h$ is taken to be equal to $m_\phi$.
The factors $\int {\cal L}_{7{\rm TeV}} dt$ and
$\int {\cal L}_{8{\rm TeV}} dt$ are the integrated luminosities
at the LHC center-of-mass energy $\sqrt{s}=7$ TeV and 8 TeV,
and in the case of Fig.~5 of Ref.~\cite{CMS-PAS-HIG-13-002}, they are
5.1 ${\rm fb}^{-1}$ and 19.6 ${\rm fb}^{-1}$, respectively.
$\sigma(pp \to \phi X ; 7(8) {\rm TeV})$ is the radion production cross
section at $\sqrt{s}=7(8)$ TeV, and similarly for the Higgs boson
production cross sections.

\begin{figure}[th]
\begin{center}
\includegraphics[scale=0.9]{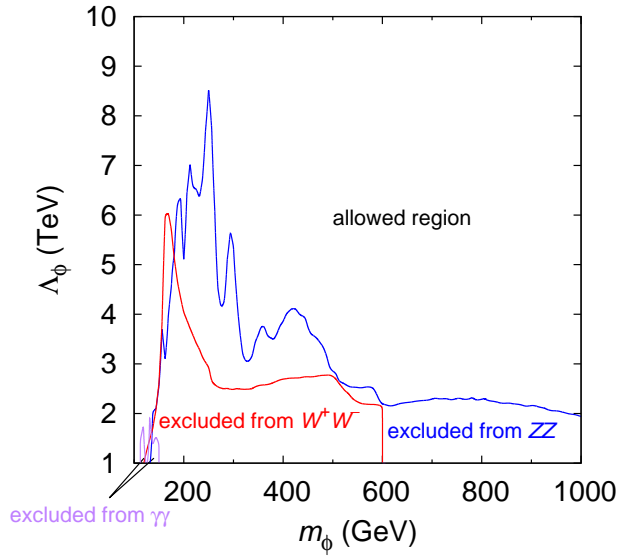}
\end{center}
\vspace*{-0.8cm}
\caption{\label{fig:ZZWWgamgam}
Excluded regions in the $(m_\phi, \Lambda_\phi)$ plane
from the SM Higgs boson searches in the $ZZ, W^+W^-$ and
$\gamma\gamma$ channels at the LHC.}
\end{figure}
In Fig.~\ref{fig:ZZWWgamgam}, we show the region in the
$(m_\phi, \Lambda_\phi)$ plane which is excluded by the
condition Eq.~(\ref{eq:the_condition}) and similar
conditions in the $W^+W^-$ and $\gamma\gamma$ channels.
The input data we use are taken from Fig.~9 in 
Ref.~\cite{CMS-PAS-HIG-13-003} for the $W^+W^-$ channel, 
the left panel of Fig.~5 in Ref.~\cite{CMS-PAS-HIG-13-002}
for the $ZZ$ channel, and Fig.~5b in Ref.~\cite{CMS-PAS-HIG-13-001}
for the $\gamma\gamma$ channel. We find that the constraint from
the $ZZ$ channel excludes wide region of the parameter space:
For $m_\phi$ around 200 GeV, the radion interaction scale $\Lambda_\phi$
is excluded up to 5 TeV.  Even for 
$m_\phi=1$ TeV,  $\Lambda_\phi$ is excluded  up to 2 TeV.
We find that this is an interesting result since the RS model
is supposed to solve the gauge hierarchy problem, and $\Lambda_{\phi}$
is believed to be not very far from the weak scale. 

Below the $VV$ $(V=W, Z)$ nominal threshold the bound
from $VV$ becomes much weaker.
In Fig.~\ref{fig:ZZWWgamgam} we find that there is a parameter region
which is excluded by $W^+W^-$ but not by $ZZ$ or $\gamma\gamma$.
Similarly, below the $W^+W^-$ nominal threshold,
there are small parameter regions which are
excluded by $\gamma\gamma$ but not by $W^+W^-$ or $ZZ$.

Before closing this section, we compare bounds on the radion parameters 
from the Higgs-boson searches and those from the first Kaluza-Klein (KK)
graviton searches. 
The mass $m_{G_1}$ of the first KK graviton in the RS model is given
by~\cite{Davoudiasl:2000wi}
\begin{eqnarray}
m_{G_1} = 3.83 k e^{-k\pi r_c}, 
\label{graviton_mass}
\end{eqnarray}
and this is related to $\Lambda_\phi$ via 
\begin{eqnarray}
 \Lambda_\phi = \sqrt{6} M_{\rm pl} e^{-k\pi r_c}, 
\label{lambda_phi_warp}
\end{eqnarray}
once the value of $k/M_{\rm pl}$ is fixed~\cite{Dominici:2002jv}.
The ATLAS collaboration has reported the lower mass bound on the first KK
graviton in the RS model as $m_{G_1}>2.23~ (1.03)~{\rm TeV}$ for 
$k/M_{\rm pl}=0.1~ (0.01)$~\cite{Aad:2012cy}, 
and this can be translated into the bound
$\Lambda_\phi > 14.3~ (65.8)$ TeV via Eqs.~(\ref{graviton_mass})
and (\ref{lambda_phi_warp}). 
Although this is a stringent bound, if the value of $k/M_{\rm pl}$
is unity, then the lower bound on $\Lambda_\phi$ is relaxed
to a few TeV~\cite{Barger:2011qn, Davoudiasl:1999jd}.
In this case, our study presented in this paper remains interesting.

\section{Summary and Discussions}

In this paper, we have studied constraints on the radion
mass and couplings in the RS model
from the LHC data on the SM Higgs boson searches.
We have used the data for $h \to ZZ$, $h \to W^+W^-$ and
$h \to \gamma\gamma$ searches which are
also useful to constrain the radion with a sub-TeV to TeV mass.
We find that the $ZZ$ data exclude wide parameter region in the
$(m_\phi, \Lambda_\phi)$ plane: for $m_\phi \sim 200$ GeV,
$\Lambda_\phi$ is excluded up to 5 TeV.   Even for
$m_\phi = 1$ TeV, the lower bound of $\Lambda_\phi$ is 2 TeV.
The $W^+W^-$ and $\gamma\gamma$ data also exclude part
of the $(m_\phi, \Lambda_\phi)$ plane, in particular, below
the nominal thresholds of $ZZ$ and $W^+W^-$, respectively. 
Since the original motivation of the RS model is to solve the
gauge hierarchy problem of the SM, the scale $\Lambda_\phi$
must not be very far from the weak scale.  In view of this,
we find that our results are interesting.  We believe that 
further experimental studies in these channels at the LHC will
give stronger bound or may find a signal of the radion.


Some comments are in order.  

Firstly, in this paper we have not considered possible mixing
between the Higgs boson and the radion.  As already mentioned,
this is because the observed Higgs boson at the LHC has 
signal strengths which are very similar to those
expected for the SM Higgs boson and hence naively we expect
the mixing between the Higgs boson and the radion to be small.
Strictly speaking, this logic is too loose, and there may
still be a parameter space in which the observed Higgs boson
is an admixture of the Higgs boson and the radion.  Although
such a study is beyond the scope of this paper, it
will be performed in a forthcoming paper~\cite{ChoNomura}.  

Secondly, in this paper we have studied only the
case where the SM fields are localized on the visible
brane.  It is clearly desirable to relax this assumption
and allow the SM fields to propagate into the bulk
since such a class of models is less constrained
from flavor physics.  

Thirdly, it is known in literature that if one assumes
the Goldberger-Wise mechanism to generate the radion mass,
then the couplings between the radion and the massless
SM gauge bosons could receive sizable corrections if the scale
$\Lambda_\phi$ is not very large compared to the radion
mass~\cite{Chacko_etal}.

A study which takes into account 
the above three issues will be published
elsewhere~\cite{ChoNomura}.  

\vspace{1cm}
\noindent
{\bf\large Acknowledgments}

\vspace{4mm}
\noindent
Y.\ O.\ would like to thank Sho Iwamoto for discussions and suggestions.
The work of G.C.C is supported in part by Grants-in-Aid for Scientific 
Research from the Ministry of
Education, Culture, Sports, Science and Technology (No.24104502) and 
from the Japan Society for the Promotion of Science (No.21244036).  

\end{document}